\documentclass[english,prb,aps,twocolumn,10pt,floatfix,superscriptaddress]{revtex4-2}

\usepackage{amsmath}
\usepackage{amssymb}
\usepackage{wasysym}
\usepackage{graphicx}
\usepackage[english]{babel}
\usepackage{array}
\usepackage{verbatim}
\usepackage[colorlinks=true, pdfstartview=FitV, linkcolor=blue, citecolor=blue, urlcolor=blue]{hyperref} %
\usepackage{xcolor}
\usepackage{braket}

\def\ie       {{\it i.e.}}

\newcommand{\ii}{\mathrm{i}}
\newcommand{\mr}[1]{\mathrm{#1}}
\newcommand{\unit}[1]{\,\mathrm{#1}}

\newcommand{\BW}{\Omega_\text{BW}}

\newcommand{\dd}{\mathrm{d}}
\newcommand{\mS}{m_S}
\newcommand{\dphi}{\delta\phi}
\newcommand{\phieff}{\phi_\mr{eff}}
\newcommand{\RR}{\hat{R}}
\newcommand{\RRd}{\hat{R}'}
\newcommand{\RXa}{\hat{R}_{\pm X}^\alpha}
\newcommand{\RYa}{\hat{R}_{Y}^\alpha}
\newcommand{\Rphi}{\hat{R}^\phi}

\newcommand{\Dt}{\Delta t}

\newcommand{\tmin}{t_\mr{min}}

\newcommand{\tML}{t_\mathrm{QSL}^{(\mathrm{ML})}}
\newcommand{\tMT}{t_\mathrm{QSL}^{(\mathrm{MT})}}
\newcommand{\tRR}{t_\mr{R}}
\newcommand{\tRRd}{t_\mr{R'}}
\newcommand{\dw}{\delta\omega}
\newcommand{\wo}{\omega_0}

\begin{document}
	
\title{Quantum speed limit in quantum sensing}
\author{Konstantin Herb}
\email{science@rashbw.de}
\affiliation{Department of Physics, ETH Z\"urich, 8093 Z\"urich, Switzerland.}
\author{Christian L. Degen}
\email{degenc@ethz.ch}
\affiliation{Department of Physics, ETH Z\"urich, 8093 Z\"urich, Switzerland.}
\affiliation{Quantum Center, ETH Z\"urich, 8093 Z\"urich, Switzerland.}
	
\begin{abstract}
	Quantum sensors capitalize on advanced control sequences for maximizing sensitivity and precision. However, protocols are not usually optimized for temporal resolution. Here, we establish the limits for time-resolved sensing of dynamical signals using qubit probes. We show that the best possible time resolution is closely related to the quantum speed limit (QSL), which describes the minimum time needed to transform between basis states. We further show that a composite control sequence consisting of two phase-shifted pulses reaches the QSL. Practical implementation is discussed based on the example of the spin-1 qutrit of a nitrogen-vacancy (NV) center in diamond.
\end{abstract}
	 
\date{\today}
\maketitle

The energy-time uncertainty principle introduced by Heisenberg is a fundamental concept of quantum mechanics.  While formulated loosely in Heisenberg's initial work, Robertson~\cite{robertson29} and Bohr~\cite{bohr28} put it on firm ground by formalizing the relationship between uncertainty and non-commutativity of observables.  Twenty years later, Mandelstam and Tamm~\cite{mandelstam45} showed that the energy-time uncertainty is not an uncertainty relation due to non-commutativity, but rather a statement about intrinsic time scales of quantum systems~\cite{deffner17}.  This insight led to the derivation of a quantum speed limit (QSL).  To do so, Mandelstam and Tamm used the von Neumann equation with the projection operator to develop an expression for the overlap between the initial state $\ket{\psi(0)}$ and the time-evolved state $\ket{\psi(t)}$, yielding
\begin{align}
	\braket{\psi(0)|\psi(t)} \geq \cos \left(\frac{\langle \Delta H\rangle t}{\hbar}\right) \ ,
	\label{eq:orthogonal}
\end{align}
in the domain $0\leq t\leq \tMT$ where $\langle \Delta H\rangle^2 =\langle H^2 \rangle - \langle H \rangle^2 $ is the variance of the Hamiltonian.  The minimum time needed to obtain a fully orthogonal state, \ie, $\braket{\psi(0)|\psi(t)}=0$, is then given by
\begin{align}
	t \geq \tMT = \frac{\pi}{2}\frac{\hbar}{\langle \Delta H \rangle} \ .
	\label{eq:qslmt}
\end{align}
Later, Margolus and Levitin~\cite{margolus98} proposed an alternative route to deriving a QSL based on the integrability of the Schr\"odinger equation to obtain a maximum evolution speed, resulting in
\begin{align}
	t \geq \tML = \frac{\pi}{2}\frac{\hbar}{\langle H \rangle} \ .
	\label{eq:qslml}
\end{align}
In contrast to the Mandelstam-Tamm definition, Eq.~\eqref{eq:qslml} bounds the time on the mean energy $\langle H\rangle$ defined relative to the energy of the ground state.  For a two-level system, the two QSLs coincide while for systems open to a continuum of states or for systems consisting of more than two states, such as qutrits~\cite{ness22}, the unified bound is tight~\cite{levitin09,ness21}.

Since the initial formulations of the QSL, theoretical work has focused on more complex situations such as open quantum systems~\cite{zhang14srep} and systems with quantum entanglement~\cite{giovannetti03}, as well as applications in quantum information processing~\cite{Svozil05}. Here, the QSL sets the maximum speed at which computations can be performed and therefore permits deriving an upper bound of the computational limits of the universe~\cite{Lloyd00,Lloyd02}. Further, the QSL has been studied in the field of quantum optimal control~\cite{Caneva09,Basilewitsch20}, quantum thermodynamics~\cite{Campaioli17}, and metrology with respect to quantum clocks~\cite{Woods22}.  Obviously, the finite time of qubit operations will also limit the time resolution achievable in quantum sensing tasks~\cite{degen17}. The time limitation is of practical importance because of the many existing and envisioned applications of quantum sensors.

In this work, we investigate the relation between the QSL and the time resolution achievable in quantum sensing experiments.  We show that a composite pulse sequence consisting of two phase-shifted control rotations, equivalent to a Ramsey sequence with zero time delay, reaches the QSL.  Opposite to a Ramsey interferometry measurement, however, phase accumulation occurs \textit{during} control rotations rather than a free evolution interval.  We derive quantitative expressions for the quantum phase pick-up as a function of control rotation angle and velocity.  We use these expressions to define the time resolution and bandwidth of the sensing sequence.  We also show that time resolution can be extended beyond the QSL by trading for a reduced signal-to-noise ratio. As an example, we simulate the expected response for a nitrogen-vacancy (NV) center in diamond that is exposed to rapidly varying magnetic signal.

\begin{figure}
\includegraphics[width=0.99\columnwidth]{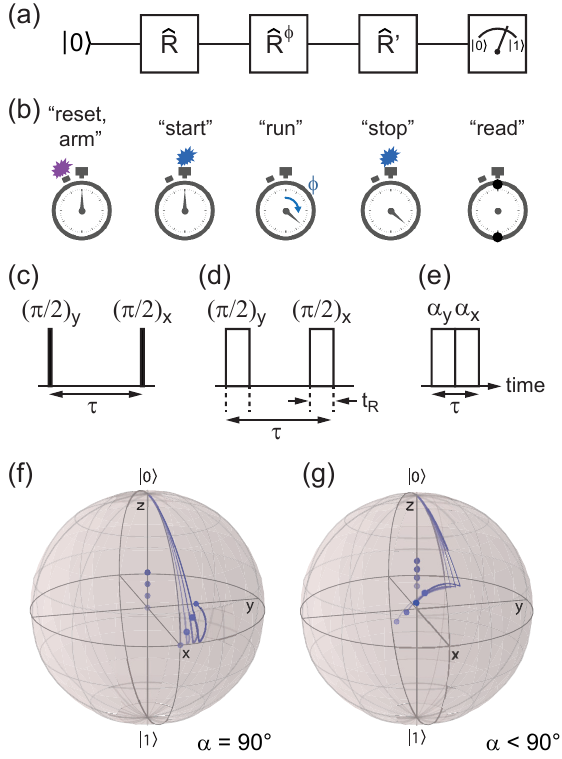}
\caption{ 
{Concept of time-resolved sensing at the QSL.}
(a) Canonical control sequence for signal estimation via quantum phase accumulation. $\RR$ and $\RRd$ are control rotations, and $\Rphi$ is the qubit rotation due to interaction with the signal.
(b) Stopwatch analogy of (a).
(c) Pulse-timing diagram of the control sequence for infinitely fast control rotations.  Phase accumulation occurs entirely during the free evolution time $\tau$.  $\pi/2$ are rotation angles and $x$, $-y$ are axes of rotation.
(d) Pulse-timing diagram for finite duration $\tRR>0$ of control rotations.
(e) Pulse-timing diagram at the QSL, when the interpulse delay becomes zero and $\tau = 2\tRR$. Phase accumulation now occurs entirely \textit{during} control rotations. $\alpha$ is the rotation angle.
(f) Bloch sphere trajectories for sequence (e) with $\alpha=90^\circ$. %
Dots are the projection on the $Z$ axis.
Trajectories are shown for $\phi = 0^\circ$ (light blue), $10^\circ$, $20^\circ$ and $30^\circ$ (dark blue).
(g), Bloch sphere trajectories for $\alpha = 63^\circ$. Further trajectories are shown in~\cite{supplemental}.
}
\label{fig1}
\end{figure}

The canonical quantum sensing scheme of Ramsey interferometry is shown in Fig.~\ref{fig1}(a)~\cite{taylor08,degen17}.  In its most basic form, the scheme uses two state transformations to initiate and halt the coherent evolution of a probe qubit subject to an external signal.  The first transformation rotates the qubit from a known initial basis state $\ket{\psi(0)}=\ket{0}$ into a coherent superposition state $\frac{1}{\sqrt{2}}(\ket{0}+\ket{1})$, which then evolves for a given time $\tau$ into state $\frac{1}{\sqrt{2}}(\ket{0}+e^{-\ii\phi(\tau)}\ket{1})$, thereby acquiring a quantum phase
\begin{align}
	\phi = \int_{0}^{\tau}[\wo+\dw(t')]\,\dd t' \ .
	\label{eq:phi}
\end{align}
Here, $\hbar\wo$ is the static energy gap between the qubit's energy levels, and $\hbar\dw(t)$ accounts for a small, time-dependent modulation due to the presence of the signal.  After coherent evolution, a second state transformation rotates the qubit back to the original basis, followed by projective state readout~\footnote{In general, the initial and readout bases may be different, but for simplicity and without loss of generality, we assume them to be the same~\cite{degen17}}. By this, the canonical quantum sensing experiment measures the expectation value of the projector onto the initial state $\ket{0}$. 
The state transformations can be described by two control rotations, $\RR$ and $\RRd$, respectively. These are analogous to the start and stop triggers in a classical test and measurement task (Fig.~\ref{fig1}(b)).  There, the time elapsed between start and stop events determines the time resolution.

Ideally, the control rotations $\RR$ and $\RRd$ are infinitely fast [Fig.~\ref{fig1}(c)].
However, owing to the QSL, $\RR$ and $\RRd$ must have a finite duration $\tRR$ or, equivalently, a finite angular velocity (Rabi frequency $\Omega$). In experiments, the maximum $\Omega$ may be limited by many factors, including finite available driving power, competition between $\omega_0$ and $\Omega$, excitation of further lying energy levels, or a combination of those. 
The finite speed of control rotations fundamentally limits the time resolution of the sensing sequence.

For finite $\tRR$, qubit evolution under the signal Hamiltonian already occurs during the control rotations.  The qubit-acquired phase then becomes the sum of the phase pick-up during the free evolution interval and the phase pick-up during control rotations [Fig.~\ref{fig1}(d)].  In the most extreme case, the free evolution time is zero, and phase pick-up occurs entirely during control rotations [Fig.~\ref{fig1}(e)].  In this situation, the time resolution of the sequence reaches that of the QSL defined by Eqs.~(\ref{eq:qslmt},\ref{eq:qslml}).  (Note that for a qubit system, the two definitions are equivalent.)

We next derive quantitative relationships between the phase pick-up $\phieff$, the velocity $\Omega$ and the angle $\alpha$ of control rotations.  For our derivation, we focus on a bipartite control sequence consisting of two consecutive qubit rotations $\RR=\RYa$ and $\RRd=\RXa$ of equal duration $\tRR$ [Fig.~\ref{fig1}(e)].  Here, $X$ and $Y$ are orthogonal axes, as shown in Fig.~\ref{fig1}(f).  This sequence represents the basic Ramsey scheme with zero time delay.  The sequence duration is $\tau = 2\tRR = \alpha/\Omega$, where $\Omega$ is the maximum allowed rotation velocity.
In the Supplemental Material~\cite{supplemental}, we show that an equal time-share ($\tRR=\tRRd$) between the two parts of the bipartite pulse sequence yields optimum sensitivity (as defined by Eq.~\eqref{eq:sensitivity} below). This is also true when varying the angle between rotation axes.

Our sensor output is the overlap between the initial qubit state $\ket{\psi(0)} = \ket{0}$ and the final qubit state $\ket{\psi(\tau)}$, given by the transition probability~\cite{degen17,giovannetti03}
\begin{subequations}
\begin{align}
	p &= 1 - |\braket{0|\psi(\tau)}|^2 \label{eq:p} \\
  	&= 1 - |\bra{0}\RR \RRd \ket{0}|^2 \label{eq:pRR} \\
	&= 1 - |\bra{0}\RXa\RYa \ket{0}|^2 .
\end{align}
\end{subequations}
In the limit of strong control fields ($\Omega\gg\dw(t)$), within linear response ($\phi\ll\pi/2$), and assuming that $\dw(t)\approx \dw$ is stationary during $\tau$, the transition probability of this sequence is~\cite{supplemental}
\begin{align}
    p &= \underbrace{\frac{1}{4}(1-\cos 2\alpha)}_{p_0}
       + \underbrace{\frac{1}{2}\frac{\sin\alpha(\cos\alpha-1)}{\alpha}\phi}_{\delta_p} \ .
	\label{eq:pexplicit}
\end{align}
Here, $p_0$ is the bias point of the measurement, and $\delta p$ is the change in probability due to the presence of the signal~$\dw$\cite{degen17}.
Eq.~(\ref{eq:pexplicit}) defines an effective phase
\begin{align}
	\delta p = \frac12 \phieff = \frac12 \frac{\sin\alpha(\cos\alpha-1)}{\alpha} \phi := \frac12 \epsilon \phi
	\label{eq:phieff}
\end{align}
that corresponds to the ideal Ramsey phase $\phi = \dw\tau$ [Eq.~(\ref{eq:phi})] reduced by the scaling factor $\epsilon<1$.
Using Eq.~\eqref{eq:phieff}, we can define the sensitivity $\eta$ of the sequence by
\begin{align}
	\eta := d[\delta p] / d[\dw] \ ,
	\label{eq:sensitivity}
\end{align}
taken in the limit $\dw\rightarrow 0$.

\begin{figure}
	\includegraphics[width=0.99\columnwidth]{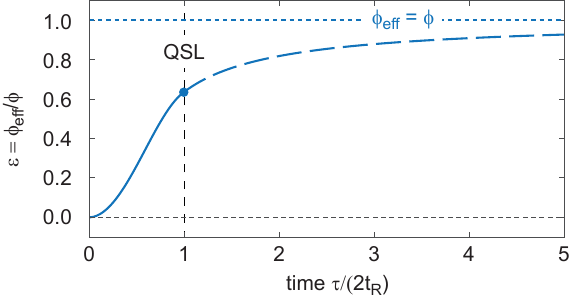}
	\caption{ 
		Phase pick-up $\phieff/\phi$ as a function of $\tau$ for control rotations with finite velocity.  For durations shorter than the QSL ($\tau<2\tRR$, solid curve), $\phieff$ is determined by the sequence of Fig.~\ref{fig1}(e).  For durations longer than the QSL ($\tau>2\tRR$, dashed curve) $\phieff$ is determined by the sequence of Fig.~\ref{fig1}(d). The upper bound ($\phieff=\phi$, dotted line) is for the hypothetical case of infinitely fast control rotations, corresponding to the sequence in Fig.~\ref{fig1}(c).
	}
	\label{fig2}
\end{figure}

For the canonical case of qubit rotations between orthogonal axes on the Bloch sphere ($\alpha = 90^\circ$), the transition probability is
\begin{align}
	p = \frac12 + \frac{\phi}{\pi}
\end{align}
corresponding to $p_0 = 0.5$ and $\epsilon = 2/\pi \approx 0.637$.  Fig.~\ref{fig1}(f,g) shows Bloch-sphere trajectories for $\alpha = 90^\circ$ and $\alpha < 90^\circ$, and Fig.~\ref{fig2} plots $\phieff$ as a function of $\tau$.

\begin{figure}
	\includegraphics[width=0.99\columnwidth]{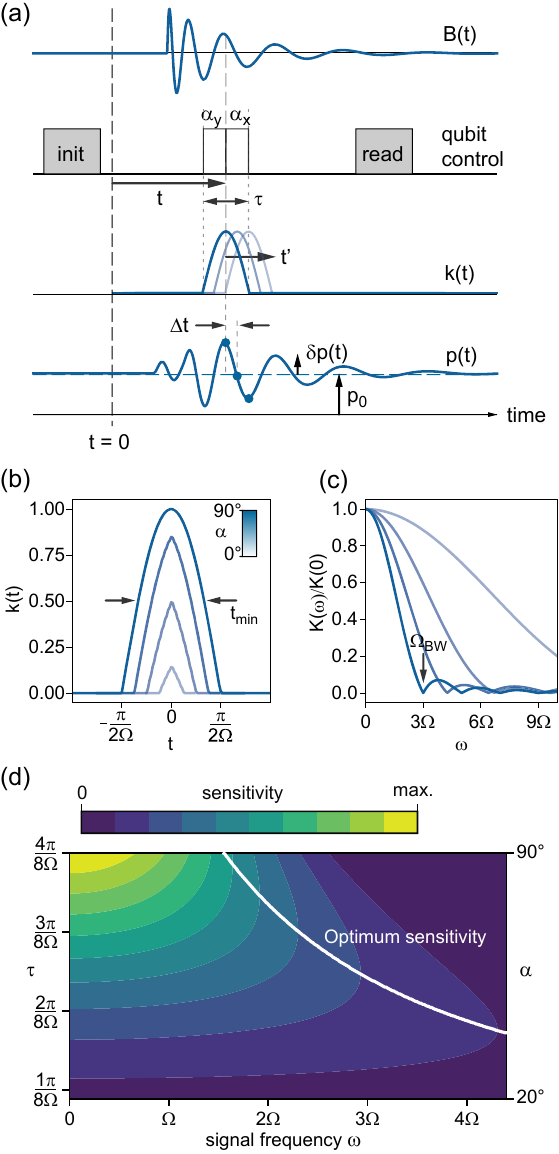}
	\caption{
		{Qubit output and sensing kernel.}
		(a) Timing diagram. To sample the signal transient $B(t)$, the control sequence (second line) is stepped along $t$ in increments of $\Dt$.  The output signal, given by state probability $p(t)$, is the convolution between $B(t)$ and the kernel $k(t)$ of the control sequence.  Gray blocks represent initialization and readout of the qubit state, and white blocks represent the control rotations.
		(b) Kernel $k(t)$ for control rotation angles $\alpha = 22.5^\circ$ (light blue), $45^\circ$, $67^\circ$ and $90^\circ$ (dark blue).  Kernels are computed using a lab-frame simulation of the spin evolution~\cite{supplemental}.  The time resolution $\tmin$ is defined by the full width at half maximum.
		(c) Normalized Bode plots $K(\omega)$ for the kernels shown in (b).  The bandwidth $\BW$ is defined by the first root of $K(\omega)$.
		(d) Sensitivity $\eta \propto K(\omega)$ as a function of signal frequency $\omega$ and pulse duration $\tau$ at fixed Rabi frequency $\Omega$.  For dynamical signals $\omega>0$, an optimum pulse duration exists where sensitivity is maximized (white curve).
	}
	\label{fig3}
\end{figure}

Having established the phase pick-up of the bipartite control sequence, we proceed to the problem of sampling a time-dependent signal.  As a generic example, we consider the detection of a transient magnetic field $\vec{B}(t)$ using a spin-1/2 system as the qubit probe where $\dw(t) = \gamma B(t)$. Here, $\gamma$ is the transduction factor (given by the gyromagnetic ratio), and $B(t)$ is the vector component of $\vec{B}(t)$ parallel to the quantization axis of the spin qubit.
The influence of off-axis components of $\vec{B}(t)$ is discussed in~\cite{supplemental}.  

To record a time transient, we sample $B(t)$ point-by-point by incrementing the delay time $t$ of the control sequence with respect to a common start trigger at $t=0$ [Fig.~\ref{fig3}(a)].  For signals that vary only slowly with time, $B(t)$ is almost stationary during control pulses, and the sensor output $\delta p(t) = \epsilon\tau\gamma B(t)$ is directly proportional to the signal field.
On the other hand, for more rapidly changing signals, the acquired quantum phase becomes a more complex function of signal and control fields.
In a general case, we express the transition probability $\delta p$ by the convolution
\begin{equation}
  \delta p(t) = \int_{t'=-\infty}^{\infty} k(t'-t) \gamma B(t') \text{d}t'
	\label{eq:convolution}
\end{equation}
where $k(t)$ is the \textit{kernel} of the control sequence and $t'$ the time delay between signal and control fields [Fig.~\ref{fig3}(a)].
Specifically, for the bipartite sequence of Fig.~\ref{fig1}(e), the sensing kernel is
\begin{align}
  k(t) =
	  \begin{cases}
			\sin[\Omega(\tau/2-|t|)] & |t| < \tau/2 \\
			0                        & |t| > \tau/2 
		\end{cases} \ .
	\label{eq:kerneltime}
\end{align}
In the frequency domain, the kernel is given by the transfer function $K(\omega) = |\mathrm{FT}[k(t)]|$,
\begin{align}
K(w) = \frac{\sqrt{\frac{2}{\pi }} \Omega  \left| \cos \left( \Omega\tau/2 \right)-\cos \left(\omega\tau/2 \right)\right| }{\left| \Omega ^2-\omega ^2\right| } \ ,
\label{eq:kernelfrequency}
\end{align}
where $\text{FT}$ is the Fourier transform.  $K(\omega)$ for other sensing sequences, such as those using amplitude-shaped pulses, can be computed numerically using spin dynamics simulations~\cite{supplemental}.
Amplitude shaping will, however, always worsen time resolution, because qubit rotations are below the maximum possible $\Omega$.

Figs.~\ref{fig3}(b,c) plots kernel profiles $k(t)$ and corresponding transfer functions $K(\omega)$ for several rotation angles $\alpha=\Omega\tau$ between $0^\circ$ and $90^\circ$.  Clearly, a smaller $\alpha$ leads to narrower kernels and, thus, an improved time resolution.
Defining the time resolution $\tmin$ by the full width at half maximum of $k(t)$, we find for the bipartite control pulse that
\begin{equation}
	\tmin :=  \tau \left(1 - \frac{\arcsin\frac{\sin\alpha}{2}}{\alpha}\right)
	\label{eq:tmin}
\end{equation}
(See~\cite{supplemental} for other possible definitions of $\tmin$).  Specifically, $\tmin = \frac23\tau = \frac{2\pi}{3\Omega}$ for $\alpha=90^\circ$, and $\tmin \approx \frac12\tau = \frac{\alpha}{2\Omega}$ in the limit $\alpha\rightarrow 0$.
Further, we can define a frequency bandwidth by the first root of the transfer function $K(\omega)$, given by
\begin{equation}
	\BW := \Omega \left(\frac{2\pi}{\alpha}-1\right),
	\label{eq:bw}
\end{equation}
where $\BW = 3\Omega$ for $\alpha=90^\circ$ (Fig.~\ref{fig3}(c) and~\cite{supplemental}).

For a given angular velocity $\Omega$, shorter rotation angles $\alpha<90^\circ$ therefore provide a means to further improve the time resolution.  The improvement is approximately $\tmin \propto \alpha$, and correspondingly for the bandwidth, $\BW \propto \alpha^{-1}$.
The improved time resolution, however, comes at the expense of a drastically lowered sensitivity, since $\eta \propto \alpha^2$ for small $\alpha$ [Eqs.~(\ref{eq:phieff},\ref{eq:sensitivity})].
The sensitivity is further illustrated in Fig.~\ref{fig3}(d), which plots $\eta$ as a function of $\tau$ for arbitrary signal frequencies $\omega$.

The apparent improvement of the time resolution beyond the QSL for short $\alpha$ [Eq.~\eqref{eq:tmin}] is in accordance with the interpretation of the uncertainty principle as defined by Eq.~\eqref{eq:orthogonal}: the QSL reflects the minimum time required for transferring a quantum state to a \emph{fully orthogonal} state. By reducing the orthogonality, thus not fully transferring the state, the time requirement shrinks therefore permitting a higher time resolution. %

\begin{figure}[t!]
	\includegraphics[width=0.99\columnwidth]{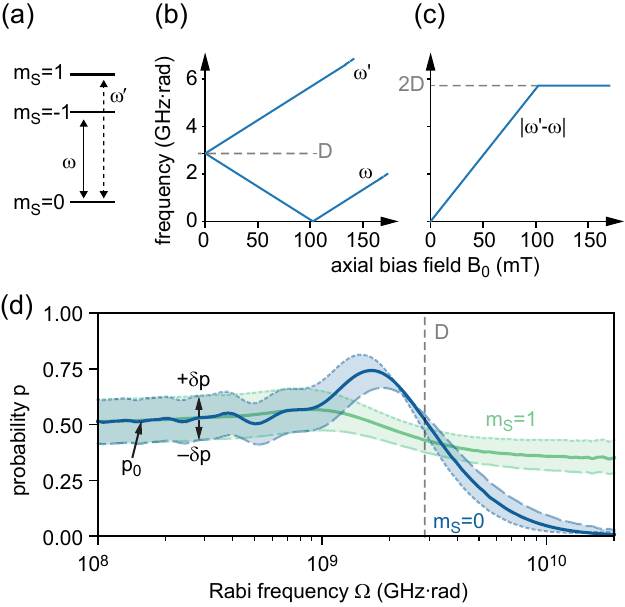}
	\caption{
		{Ultimate limits to time resolution for an NV probe qutrit.}
		(a) NV spin energy levels and allowed transitions $\omega$ and $\omega'$.
		(b) Scaling of transition frequencies $\omega$ and $\omega'$ with axial bias field $B_0$. $D = 2.87\unit{GHz}$ is the zero-field splitting parameter.
		(c) Isolation of transition frequencies $|\omega'-\omega|$ as a function of $B_0$.  The frequency difference saturates at $2D$.
		(d) Numerical simulation of the state probability $p$ as a function of driving frequency $\Omega$. $p_0$ is the bias point and $\delta p$ the probability change caused by the signal, see Eq.~(\ref{eq:pexplicit}).  At high $\Omega/2\pi>D$, the probability signal inverts and goes to zero for $\Omega\rightarrow\infty$ for preparation and measurement in the $\mS=0$ basis (blue), but not in the $\mS=\pm1$ (green) basis.  For the simulation, a fixed value of $\dphi = 0.2$ was taken independent of $\Omega$.
	}
	\label{fig4}
\end{figure}

As a practical example, we consider the $S=1$ spin system of a single nitrogen-vacancy (NV) center in diamond~\cite{doherty13}.
The NV center is a prototypical qubit sensor with a growing range of applications in materials science, physics, chemistry, and biology~\cite{schirhagl14,barry16,janitz22,petrini22,finco23,rovny24}, including the study of dynamical excitations in these systems~\cite{vandersar15,wolfe16,zhou20}.
The NV center exhibits three spin energy levels $\mS=0$ and $\mS=\pm 1$ and two allowed spin transitions with frequencies $\omega$ and $\omega'$, as depicted in Fig.~\ref{fig4}(a).  To form an effective two-level system, the $\omega$ (or $\omega'$) transition can be isolated by applying an axial bias field $B_0$ along the NV symmetry axis [Fig.~\ref{fig4}(b)].  For driving fields $\Omega$ smaller than the frequency difference $|\omega'-\omega|$ between the $\mS=\pm1$ levels [Fig.~\ref{fig4}(c)], the excitation is selective and the other ($\omega'$ or $\omega$) transition can be neglected.

Fig.~\ref{fig4}(d) shows the expected output probability $p$ for an NV spin qubit as a function of $\Omega$.
The output probability is calculated using a laboratory-frame simulation of the spin dynamics, taking the full $S=1$ nature of the NV center and the effect of counter-rotating terms in the excitation pulses into account~\cite{supplemental}.
Two simulations are presented where the spin is either initialized and read out in the $\mS=0$ state or in the $\mS=-1$ state, respectively.

First, we consider the $\mS=0$ case. 
For moderate driving fields $\Omega$ that are safely smaller than any of the transition frequencies ($\omega$, $\omega'$, $|\omega'-\omega|$), the output probability $p$ is nearly independent of $\Omega$ and follows the theory of Eq.~(\ref{eq:phieff}).  
On the other hand, for larger $\Omega$, the output probability is modified due to strong driving effects -- namely, Bloch-Siegert shifts and breakdown of the rotating-wave approximation~\cite{bloch40,fuchs09,kairys23} -- and by spurious excitation of the $\omega'$ transition.  Both effects are not accounted for by the basic theory of Eq.~(\ref{eq:phieff}).
Further, because the $\omega$ and $\omega'$ transitions lead to opposite signs in the phase $\phieff$, the signal is completely canceled for very large $\Omega \gg |\omega'-\omega|$ (blue trace in Fig.~\ref{fig4}(d)).

By contrast, when preparing and reading out the $\mS=-1$ spin state (green trace in Fig.~\ref{fig4}(d)), the cancellation remains incomplete at any bias field.  This provides a remedy to the $\mS=0$ case.  The incomplete cancellation is due to selection rules for qubit rotations: transitions are allowed between $\mS=0$ and $\mS=\pm1$ but forbidden between $\mS=-1$ and $\mS=+1$ states.  Preparation and readout of $\mS=\pm1$ are realized experimentally by slow, selective $\pi$ rotations.

In summary, our work establishes limits to the temporal resolution reachable by a quantum sensor.
We develop our discussion in the framework of the QSL, which is applied to the canonical sensing principle of Ramsey interferometry.
We derive expressions for the coherent phase pick-up during a generic sequence consisting of two phase-shifted control pulses and analyze the response regarding time resolution, frequency bandwidth, and sensitivity.
We numerically simulate the expected phase response for the single spin of an NV center in diamond, taking the full $S=1$ nature of the spin system and nonlinear driving effects into account.
Beyond fundamental aspects in quantum metrology, our work has practical relevance for real-world applications of quantum sensors, such as the mapping of fast magnetization reversals in spintronic devices.  These dynamics typically occur on time scales of a few nanoseconds~ \cite{baumgartner17}.  For NV centers, $\tmin \sim 1\unit{ns}$ is reached, for example, using $\Omega/2\pi = 100\unit{MHz}$ and $\alpha = 60^\circ$~[Eq.~(\ref{eq:tmin})].  Such Rabi frequencies $\Omega$ are available using on-chip microwave delivery~\cite{fuchs09}. An initial experimental demonstration of our technique is discussed in Ref. \cite{arxiv_nanosecond}.

The authors thank John Abendroth, Joseph Renes and Laura Alicia V\"olker for fruitful discussions.  This work has been supported by Swiss National Science Foundation (SNSF) Project Grant No. 200020\_212051/1 and by the European
Research Council through ERC CoG 817720 (IMAGINE).

\bibliography{library}

\begin{thebibliography}{36}%
\makeatletter
\providecommand \@ifxundefined [1]{%
 \@ifx{#1\undefined}
}%
\providecommand \@ifnum [1]{%
 \ifnum #1\expandafter \@firstoftwo
 \else \expandafter \@secondoftwo
 \fi
}%
\providecommand \@ifx [1]{%
 \ifx #1\expandafter \@firstoftwo
 \else \expandafter \@secondoftwo
 \fi
}%
\providecommand \natexlab [1]{#1}%
\providecommand \enquote  [1]{``#1''}%
\providecommand \bibnamefont  [1]{#1}%
\providecommand \bibfnamefont [1]{#1}%
\providecommand \citenamefont [1]{#1}%
\providecommand \href@noop [0]{\@secondoftwo}%
\providecommand \href [0]{\begingroup \@sanitize@url \@href}%
\providecommand \@href[1]{\@@startlink{#1}\@@href}%
\providecommand \@@href[1]{\endgroup#1\@@endlink}%
\providecommand \@sanitize@url [0]{\catcode `\\12\catcode `\$12\catcode
  `\&12\catcode `\#12\catcode `\^12\catcode `\_12\catcode `\%12\relax}%
\providecommand \@@startlink[1]{}%
\providecommand \@@endlink[0]{}%
\providecommand \url  [0]{\begingroup\@sanitize@url \@url }%
\providecommand \@url [1]{\endgroup\@href {#1}{\urlprefix }}%
\providecommand \urlprefix  [0]{URL }%
\providecommand \Eprint [0]{\href }%
\providecommand \doibase [0]{https://doi.org/}%
\providecommand \selectlanguage [0]{\@gobble}%
\providecommand \bibinfo  [0]{\@secondoftwo}%
\providecommand \bibfield  [0]{\@secondoftwo}%
\providecommand \translation [1]{[#1]}%
\providecommand \BibitemOpen [0]{}%
\providecommand \bibitemStop [0]{}%
\providecommand \bibitemNoStop [0]{.\EOS\space}%
\providecommand \EOS [0]{\spacefactor3000\relax}%
\providecommand \BibitemShut  [1]{\csname bibitem#1\endcsname}%
\let\auto@bib@innerbib\@empty
\bibitem [{\citenamefont {Robertson}(1929)}]{robertson29}%
  \BibitemOpen
  \bibfield  {author} {\bibinfo {author} {\bibfnamefont {H.~P.}\ \bibnamefont
  {Robertson}},\ }\bibfield  {title} {\bibinfo {title} {The uncertainty
  principle},\ }\href {https://doi.org/10.1103/PhysRev.34.163} {\bibfield
  {journal} {\bibinfo  {journal} {Phys. Rev.}\ }\textbf {\bibinfo {volume}
  {34}},\ \bibinfo {pages} {163} (\bibinfo {year} {1929})}\BibitemShut
  {NoStop}%
\bibitem [{\citenamefont {Bohr}(1928)}]{bohr28}%
  \BibitemOpen
  \bibfield  {author} {\bibinfo {author} {\bibfnamefont {N.}~\bibnamefont
  {Bohr}},\ }\bibfield  {title} {\bibinfo {title} {The quantum postulate and
  the recent development of atomic theory},\ }\href
  {https://doi.org/10.1038/121580a0} {\bibfield  {journal} {\bibinfo  {journal}
  {Nature}\ }\textbf {\bibinfo {volume} {121}},\ \bibinfo {pages} {580–590}
  (\bibinfo {year} {1928})}\BibitemShut {NoStop}%
\bibitem [{\citenamefont {Mandelstam}(1945)}]{mandelstam45}%
  \BibitemOpen
  \bibfield  {author} {\bibinfo {author} {\bibfnamefont {L.}~\bibnamefont
  {Mandelstam}},\ }\bibfield  {title} {\bibinfo {title} {The uncertainty
  relation between energy and time in nonrelativistic quantum mechanics},\
  }\href {https://cir.nii.ac.jp/crid/1370285709897301766} {\bibfield  {journal}
  {\bibinfo  {journal} {J. Phys. (USSR}\ }\textbf {\bibinfo {volume} {9}},\
  \bibinfo {pages} {249} (\bibinfo {year} {1945})}\BibitemShut {NoStop}%
\bibitem [{\citenamefont {Deffner}\ and\ \citenamefont
  {Campbell}(2017)}]{deffner17}%
  \BibitemOpen
  \bibfield  {author} {\bibinfo {author} {\bibfnamefont {S.}~\bibnamefont
  {Deffner}}\ and\ \bibinfo {author} {\bibfnamefont {S.}~\bibnamefont
  {Campbell}},\ }\bibfield  {title} {\bibinfo {title} {Quantum speed limits:
  from heisenberg’s uncertainty principle to optimal quantum control},\
  }\href {https://doi.org/10.1088/1751-8121/aa86c6} {\bibfield  {journal}
  {\bibinfo  {journal} {Journal of Physics A: Mathematical and Theoretical}\
  }\textbf {\bibinfo {volume} {50}},\ \bibinfo {pages} {453001} (\bibinfo
  {year} {2017})}\BibitemShut {NoStop}%
\bibitem [{\citenamefont {Margolus}\ and\ \citenamefont
  {Levitin}(1998)}]{margolus98}%
  \BibitemOpen
  \bibfield  {author} {\bibinfo {author} {\bibfnamefont {N.}~\bibnamefont
  {Margolus}}\ and\ \bibinfo {author} {\bibfnamefont {L.~B.}\ \bibnamefont
  {Levitin}},\ }\bibfield  {title} {\bibinfo {title} {The maximum speed of
  dynamical evolution},\ }\href
  {https://www.sciencedirect.com/science/article/pii/S0167278998000542}
  {\bibfield  {journal} {\bibinfo  {journal} {Physica D: Nonlinear Phenomena}\
  }\textbf {\bibinfo {volume} {120}},\ \bibinfo {pages} {188} (\bibinfo {year}
  {1998})}\BibitemShut {NoStop}%
\bibitem [{\citenamefont {Ness}\ \emph {et~al.}(2022)\citenamefont {Ness},
  \citenamefont {Alberti},\ and\ \citenamefont {Sagi}}]{ness22}%
  \BibitemOpen
  \bibfield  {author} {\bibinfo {author} {\bibfnamefont {G.}~\bibnamefont
  {Ness}}, \bibinfo {author} {\bibfnamefont {A.}~\bibnamefont {Alberti}},\ and\
  \bibinfo {author} {\bibfnamefont {Y.}~\bibnamefont {Sagi}},\ }\bibfield
  {title} {\bibinfo {title} {Quantum speed limit for states with a bounded
  energy spectrum},\ }\href {https://doi.org/10.1103/PhysRevLett.129.140403}
  {\bibfield  {journal} {\bibinfo  {journal} {Phys. Rev. Lett.}\ }\textbf
  {\bibinfo {volume} {129}},\ \bibinfo {pages} {140403} (\bibinfo {year}
  {2022})}\BibitemShut {NoStop}%
\bibitem [{\citenamefont {Levitin}\ and\ \citenamefont
  {Toffoli}(2009)}]{levitin09}%
  \BibitemOpen
  \bibfield  {author} {\bibinfo {author} {\bibfnamefont {L.~B.}\ \bibnamefont
  {Levitin}}\ and\ \bibinfo {author} {\bibfnamefont {T.}~\bibnamefont
  {Toffoli}},\ }\bibfield  {title} {\bibinfo {title} {Fundamental limit on the
  rate of quantum dynamics: The unified bound is tight},\ }\href
  {https://doi.org/10.1103/PhysRevLett.103.160502} {\bibfield  {journal}
  {\bibinfo  {journal} {Phys. Rev. Lett.}\ }\textbf {\bibinfo {volume} {103}},\
  \bibinfo {pages} {160502} (\bibinfo {year} {2009})}\BibitemShut {NoStop}%
\bibitem [{\citenamefont {Ness}\ \emph {et~al.}(2021)\citenamefont {Ness},
  \citenamefont {Lam}, \citenamefont {Alt}, \citenamefont {Meschede},
  \citenamefont {Sagi},\ and\ \citenamefont {Alberti}}]{ness21}%
  \BibitemOpen
  \bibfield  {author} {\bibinfo {author} {\bibfnamefont {G.}~\bibnamefont
  {Ness}}, \bibinfo {author} {\bibfnamefont {M.~R.}\ \bibnamefont {Lam}},
  \bibinfo {author} {\bibfnamefont {W.}~\bibnamefont {Alt}}, \bibinfo {author}
  {\bibfnamefont {D.}~\bibnamefont {Meschede}}, \bibinfo {author}
  {\bibfnamefont {Y.}~\bibnamefont {Sagi}},\ and\ \bibinfo {author}
  {\bibfnamefont {A.}~\bibnamefont {Alberti}},\ }\bibfield  {title} {\bibinfo
  {title} {Observing crossover between quantum speed limits},\ }\bibfield
  {journal} {\bibinfo  {journal} {Science Advances}\ }\textbf {\bibinfo
  {volume} {7}},\ \href {https://doi.org/10.1126/sciadv.abj9119}
  {10.1126/sciadv.abj9119} (\bibinfo {year} {2021})\BibitemShut {NoStop}%
\bibitem [{\citenamefont {Zhang}\ \emph {et~al.}(2014)\citenamefont {Zhang},
  \citenamefont {Han}, \citenamefont {Xia}, \citenamefont {Cao},\ and\
  \citenamefont {Fan}}]{zhang14srep}%
  \BibitemOpen
  \bibfield  {author} {\bibinfo {author} {\bibfnamefont {Y.}~\bibnamefont
  {Zhang}}, \bibinfo {author} {\bibfnamefont {W.}~\bibnamefont {Han}}, \bibinfo
  {author} {\bibfnamefont {Y.}~\bibnamefont {Xia}}, \bibinfo {author}
  {\bibfnamefont {J.}~\bibnamefont {Cao}},\ and\ \bibinfo {author}
  {\bibfnamefont {H.}~\bibnamefont {Fan}},\ }\bibfield  {title} {\bibinfo
  {title} {Quantum speed limit for arbitrary initial states},\ }\href
  {https://doi.org/10.1038/srep04890} {\bibfield  {journal} {\bibinfo
  {journal} {Scientific Reports}\ }\textbf {\bibinfo {volume} {4}},\ \bibinfo
  {pages} {4890} (\bibinfo {year} {2014})}\BibitemShut {NoStop}%
\bibitem [{\citenamefont {Giovannetti}\ \emph {et~al.}(2003)\citenamefont
  {Giovannetti}, \citenamefont {Lloyd},\ and\ \citenamefont
  {Maccone}}]{giovannetti03}%
  \BibitemOpen
  \bibfield  {author} {\bibinfo {author} {\bibfnamefont {V.}~\bibnamefont
  {Giovannetti}}, \bibinfo {author} {\bibfnamefont {S.}~\bibnamefont {Lloyd}},\
  and\ \bibinfo {author} {\bibfnamefont {L.}~\bibnamefont {Maccone}},\
  }\bibfield  {title} {\bibinfo {title} {Quantum limits to dynamical
  evolution},\ }\href {https://doi.org/10.1103/PhysRevA.67.052109} {\bibfield
  {journal} {\bibinfo  {journal} {Physical Review A}\ }\textbf {\bibinfo
  {volume} {67}},\ \bibinfo {pages} {052109} (\bibinfo {year}
  {2003})}\BibitemShut {NoStop}%
\bibitem [{\citenamefont {Svozil}\ \emph {et~al.}(2005)\citenamefont {Svozil},
  \citenamefont {Levitin}, \citenamefont {Toffoli},\ and\ \citenamefont
  {Walton}}]{Svozil05}%
  \BibitemOpen
  \bibfield  {author} {\bibinfo {author} {\bibfnamefont {K.}~\bibnamefont
  {Svozil}}, \bibinfo {author} {\bibfnamefont {L.~B.}\ \bibnamefont {Levitin}},
  \bibinfo {author} {\bibfnamefont {T.}~\bibnamefont {Toffoli}},\ and\ \bibinfo
  {author} {\bibfnamefont {Z.}~\bibnamefont {Walton}},\ }\bibfield  {title}
  {\bibinfo {title} {Maximum speed of quantum gate operation},\ }\href
  {https://doi.org/10.1007/s10773-005-7073-8} {\bibfield  {journal} {\bibinfo
  {journal} {International Journal of Theoretical Physics}\ }\textbf {\bibinfo
  {volume} {44}},\ \bibinfo {pages} {965–970} (\bibinfo {year}
  {2005})}\BibitemShut {NoStop}%
\bibitem [{\citenamefont {Lloyd}(2000)}]{Lloyd00}%
  \BibitemOpen
  \bibfield  {author} {\bibinfo {author} {\bibfnamefont {S.}~\bibnamefont
  {Lloyd}},\ }\bibfield  {title} {\bibinfo {title} {Ultimate physical limits to
  computation},\ }\href {https://doi.org/10.1038/35023282} {\bibfield
  {journal} {\bibinfo  {journal} {Nature}\ }\textbf {\bibinfo {volume} {406}},\
  \bibinfo {pages} {1047–1054} (\bibinfo {year} {2000})}\BibitemShut
  {NoStop}%
\bibitem [{\citenamefont {Lloyd}(2002)}]{Lloyd02}%
  \BibitemOpen
  \bibfield  {author} {\bibinfo {author} {\bibfnamefont {S.}~\bibnamefont
  {Lloyd}},\ }\bibfield  {title} {\bibinfo {title} {Computational capacity of
  the universe},\ }\href {https://doi.org/10.1103/PhysRevLett.88.237901}
  {\bibfield  {journal} {\bibinfo  {journal} {Phys. Rev. Lett.}\ }\textbf
  {\bibinfo {volume} {88}},\ \bibinfo {pages} {237901} (\bibinfo {year}
  {2002})}\BibitemShut {NoStop}%
\bibitem [{\citenamefont {Caneva}\ \emph {et~al.}(2009)\citenamefont {Caneva},
  \citenamefont {T.}, \citenamefont {Calarco}, \citenamefont {Fazio},
  \citenamefont {Montangero}, \citenamefont {Giovannetti},\ and\ \citenamefont
  {Santoro}}]{Caneva09}%
  \BibitemOpen
  \bibfield  {author} {\bibinfo {author} {\bibfnamefont {T.}~\bibnamefont
  {Caneva}}, \bibinfo {author} {\bibfnamefont {M.~M.}\ \bibnamefont {T.}},
  \bibinfo {author} {\bibnamefont {Calarco}}, \bibinfo {author} {\bibfnamefont
  {R.}~\bibnamefont {Fazio}}, \bibinfo {author} {\bibfnamefont
  {S.}~\bibnamefont {Montangero}}, \bibinfo {author} {\bibfnamefont
  {V.}~\bibnamefont {Giovannetti}},\ and\ \bibinfo {author} {\bibfnamefont
  {G.~E.}\ \bibnamefont {Santoro}},\ }\bibfield  {title} {\bibinfo {title}
  {Optimal control at the quantum speed limit},\ }\href
  {https://doi.org/10.1103/PhysRevLett.103.240501} {\bibfield  {journal}
  {\bibinfo  {journal} {Phys. Rev. Lett.}\ }\textbf {\bibinfo {volume} {103}},\
  \bibinfo {pages} {240501} (\bibinfo {year} {2009})}\BibitemShut {NoStop}%
\bibitem [{\citenamefont {Basilewitsch}\ \emph {et~al.}(2020)\citenamefont
  {Basilewitsch}, \citenamefont {Yuan},\ and\ \citenamefont
  {Koch}}]{Basilewitsch20}%
  \BibitemOpen
  \bibfield  {author} {\bibinfo {author} {\bibfnamefont {D.}~\bibnamefont
  {Basilewitsch}}, \bibinfo {author} {\bibfnamefont {H.}~\bibnamefont {Yuan}},\
  and\ \bibinfo {author} {\bibfnamefont {C.~P.}\ \bibnamefont {Koch}},\
  }\bibfield  {title} {\bibinfo {title} {Optimally controlled quantum
  discrimination and estimation},\ }\href
  {https://doi.org/10.1103/PhysRevResearch.2.033396} {\bibfield  {journal}
  {\bibinfo  {journal} {Phys. Rev. Res.}\ }\textbf {\bibinfo {volume} {2}},\
  \bibinfo {pages} {033396} (\bibinfo {year} {2020})}\BibitemShut {NoStop}%
\bibitem [{\citenamefont {Campaioli}\ \emph {et~al.}(2017)\citenamefont
  {Campaioli}, \citenamefont {Pollock}, \citenamefont {Binder}, \citenamefont
  {C\'eleri}, \citenamefont {Goold}, \citenamefont {Vinjanampathy},\ and\
  \citenamefont {Modi}}]{Campaioli17}%
  \BibitemOpen
  \bibfield  {author} {\bibinfo {author} {\bibfnamefont {F.}~\bibnamefont
  {Campaioli}}, \bibinfo {author} {\bibfnamefont {F.~A.}\ \bibnamefont
  {Pollock}}, \bibinfo {author} {\bibfnamefont {F.~C.}\ \bibnamefont {Binder}},
  \bibinfo {author} {\bibfnamefont {L.}~\bibnamefont {C\'eleri}}, \bibinfo
  {author} {\bibfnamefont {J.}~\bibnamefont {Goold}}, \bibinfo {author}
  {\bibfnamefont {S.}~\bibnamefont {Vinjanampathy}},\ and\ \bibinfo {author}
  {\bibfnamefont {K.}~\bibnamefont {Modi}},\ }\bibfield  {title} {\bibinfo
  {title} {Enhancing the charging power of quantum batteries},\ }\href
  {https://doi.org/10.1103/PhysRevLett.118.150601} {\bibfield  {journal}
  {\bibinfo  {journal} {Phys. Rev. Lett.}\ }\textbf {\bibinfo {volume} {118}},\
  \bibinfo {pages} {150601} (\bibinfo {year} {2017})}\BibitemShut {NoStop}%
\bibitem [{\citenamefont {Woods}\ \emph {et~al.}(2022)\citenamefont {Woods},
  \citenamefont {Silva}, \citenamefont {P\"utz}, \citenamefont {Stupar},\ and\
  \citenamefont {Renner}}]{Woods22}%
  \BibitemOpen
  \bibfield  {author} {\bibinfo {author} {\bibfnamefont {M.~P.}\ \bibnamefont
  {Woods}}, \bibinfo {author} {\bibfnamefont {R.}~\bibnamefont {Silva}},
  \bibinfo {author} {\bibfnamefont {G.}~\bibnamefont {P\"utz}}, \bibinfo
  {author} {\bibfnamefont {S.}~\bibnamefont {Stupar}},\ and\ \bibinfo {author}
  {\bibfnamefont {R.}~\bibnamefont {Renner}},\ }\bibfield  {title} {\bibinfo
  {title} {Quantum clocks are more accurate than classical ones},\ }\href
  {https://doi.org/10.1103/PRXQuantum.3.010319} {\bibfield  {journal} {\bibinfo
   {journal} {PRX Quantum}\ }\textbf {\bibinfo {volume} {3}},\ \bibinfo {pages}
  {010319} (\bibinfo {year} {2022})}\BibitemShut {NoStop}%
\bibitem [{\citenamefont {Degen}\ \emph {et~al.}(2017)\citenamefont {Degen},
  \citenamefont {Reinhard},\ and\ \citenamefont {Cappellaro}}]{degen17}%
  \BibitemOpen
  \bibfield  {author} {\bibinfo {author} {\bibfnamefont {C.}~\bibnamefont
  {Degen}}, \bibinfo {author} {\bibfnamefont {F.}~\bibnamefont {Reinhard}},\
  and\ \bibinfo {author} {\bibfnamefont {P.}~\bibnamefont {Cappellaro}},\
  }\bibfield  {title} {\bibinfo {title} {Quantum sensing},\ }\href
  {https://doi.org/10.1103/RevModPhys.89.035002} {\bibfield  {journal}
  {\bibinfo  {journal} {Rev. Mod. Phys.}\ }\textbf {\bibinfo {volume} {89}},\
  \bibinfo {pages} {035002} (\bibinfo {year} {2017})}\BibitemShut {NoStop}%
\bibitem [{sup()}]{supplemental}%
  \BibitemOpen
  \href@noop {} {\bibinfo  {journal} {See Supplemental Material accompanying
  this manuscript}\ }\BibitemShut {NoStop}%
\bibitem [{\citenamefont {Taylor}\ \emph {et~al.}(2008)\citenamefont {Taylor},
  \citenamefont {Cappellaro}, \citenamefont {Childress}, \citenamefont {Jiang},
  \citenamefont {Budker}, \citenamefont {Hemmer}, \citenamefont {A.Yacoby},
  \citenamefont {Walsworth},\ and\ \citenamefont {Lukin}}]{taylor08}%
  \BibitemOpen
\bibfield  {journal} {  }\bibfield  {author} {\bibinfo {author} {\bibfnamefont
  {J.~M.}\ \bibnamefont {Taylor}}, \bibinfo {author} {\bibfnamefont
  {P.}~\bibnamefont {Cappellaro}}, \bibinfo {author} {\bibfnamefont
  {L.}~\bibnamefont {Childress}}, \bibinfo {author} {\bibfnamefont
  {L.}~\bibnamefont {Jiang}}, \bibinfo {author} {\bibfnamefont
  {D.}~\bibnamefont {Budker}}, \bibinfo {author} {\bibfnamefont {P.~R.}\
  \bibnamefont {Hemmer}}, \bibinfo {author} {\bibnamefont {A.Yacoby}}, \bibinfo
  {author} {\bibfnamefont {R.}~\bibnamefont {Walsworth}},\ and\ \bibinfo
  {author} {\bibfnamefont {M.~D.}\ \bibnamefont {Lukin}},\ }\bibfield  {title}
  {\bibinfo {title} {High-sensitivity diamond magnetometer with nanoscale
  resolution},\ }\href {https://doi.org/10.1038/nphys1075} {\bibfield
  {journal} {\bibinfo  {journal} {Nat. Phys.}\ }\textbf {\bibinfo {volume}
  {4}},\ \bibinfo {eid} {810} (\bibinfo {year} {2008})}\BibitemShut {NoStop}%
\bibitem [{Note1()}]{Note1}%
  \BibitemOpen
  \bibinfo {note} {In general, the initial and readout bases may be different,
  but for simplicity and without loss of generality, we assume them to be the
  same~\cite {degen17}}\BibitemShut {NoStop}%
\bibitem [{\citenamefont {Doherty}\ \emph {et~al.}(2013)\citenamefont
  {Doherty}, \citenamefont {Manson}, \citenamefont {Delaney}, \citenamefont
  {Jelezko}, \citenamefont {Wrachtrup},\ and\ \citenamefont
  {Hollenberg}}]{doherty13}%
  \BibitemOpen
  \bibfield  {author} {\bibinfo {author} {\bibfnamefont {M.~W.}\ \bibnamefont
  {Doherty}}, \bibinfo {author} {\bibfnamefont {N.~B.}\ \bibnamefont {Manson}},
  \bibinfo {author} {\bibfnamefont {P.}~\bibnamefont {Delaney}}, \bibinfo
  {author} {\bibfnamefont {F.}~\bibnamefont {Jelezko}}, \bibinfo {author}
  {\bibfnamefont {J.}~\bibnamefont {Wrachtrup}},\ and\ \bibinfo {author}
  {\bibfnamefont {L.~C.}\ \bibnamefont {Hollenberg}},\ }\bibfield  {title}
  {\bibinfo {title} {The nitrogen-vacancy colour centre in diamond},\ }\href
  {http://www.sciencedirect.com/science/article/pii/S0370157313000562}
  {\bibfield  {journal} {\bibinfo  {journal} {Physics Reports}\ }\textbf
  {\bibinfo {volume} {528}},\ \bibinfo {pages} {1} (\bibinfo {year}
  {2013})}\BibitemShut {NoStop}%
\bibitem [{\citenamefont {Schirhagl}\ \emph {et~al.}(2014)\citenamefont
  {Schirhagl}, \citenamefont {Chang}, \citenamefont {Loretz},\ and\
  \citenamefont {Degen}}]{schirhagl14}%
  \BibitemOpen
  \bibfield  {author} {\bibinfo {author} {\bibfnamefont {R.}~\bibnamefont
  {Schirhagl}}, \bibinfo {author} {\bibfnamefont {K.}~\bibnamefont {Chang}},
  \bibinfo {author} {\bibfnamefont {M.}~\bibnamefont {Loretz}},\ and\ \bibinfo
  {author} {\bibfnamefont {C.~L.}\ \bibnamefont {Degen}},\ }\bibfield  {title}
  {\bibinfo {title} {Nitrogen-vacancy centers in diamond: Nanoscale sensors for
  physics and biology},\ }\href
  {https://doi.org/10.1146/annurev-physchem-040513-103659} {\bibfield
  {journal} {\bibinfo  {journal} {Annu. Rev. Phys. Chem.}\ }\textbf {\bibinfo
  {volume} {65}},\ \bibinfo {pages} {83} (\bibinfo {year} {2014})}\BibitemShut
  {NoStop}%
\bibitem [{\citenamefont {Barry}\ \emph {et~al.}(2016)\citenamefont {Barry},
  \citenamefont {Turner}, \citenamefont {Schloss}, \citenamefont {Glenn},
  \citenamefont {Song}, \citenamefont {Lukin}, \citenamefont {Park},\ and\
  \citenamefont {Walsworth}}]{barry16}%
  \BibitemOpen
  \bibfield  {author} {\bibinfo {author} {\bibfnamefont {J.~F.}\ \bibnamefont
  {Barry}}, \bibinfo {author} {\bibfnamefont {M.~J.}\ \bibnamefont {Turner}},
  \bibinfo {author} {\bibfnamefont {J.~M.}\ \bibnamefont {Schloss}}, \bibinfo
  {author} {\bibfnamefont {D.~R.}\ \bibnamefont {Glenn}}, \bibinfo {author}
  {\bibfnamefont {Y.}~\bibnamefont {Song}}, \bibinfo {author} {\bibfnamefont
  {M.~D.}\ \bibnamefont {Lukin}}, \bibinfo {author} {\bibfnamefont
  {H.}~\bibnamefont {Park}},\ and\ \bibinfo {author} {\bibfnamefont {R.~L.}\
  \bibnamefont {Walsworth}},\ }\bibfield  {title} {\bibinfo {title} {Optical
  magnetic detection of single-neuron action potentials using quantum defects
  in diamond},\ }\href {https://doi.org/10.1073/pnas.1601513113} {\bibfield
  {journal} {\bibinfo  {journal} {Proc. Natl. Acad. Sci. USA}\ }\textbf
  {\bibinfo {volume} {113}},\ \bibinfo {pages} {14133} (\bibinfo {year}
  {2016})}\BibitemShut {NoStop}%
\bibitem [{\citenamefont {Janitz}\ \emph {et~al.}(2022)\citenamefont {Janitz},
  \citenamefont {Herb}, \citenamefont {Volker}, \citenamefont {Huxter},
  \citenamefont {Degen},\ and\ \citenamefont {Abendroth}}]{janitz22}%
  \BibitemOpen
  \bibfield  {author} {\bibinfo {author} {\bibfnamefont {E.}~\bibnamefont
  {Janitz}}, \bibinfo {author} {\bibfnamefont {K.}~\bibnamefont {Herb}},
  \bibinfo {author} {\bibfnamefont {L.~A.}\ \bibnamefont {Volker}}, \bibinfo
  {author} {\bibfnamefont {W.~S.}\ \bibnamefont {Huxter}}, \bibinfo {author}
  {\bibfnamefont {C.~L.}\ \bibnamefont {Degen}},\ and\ \bibinfo {author}
  {\bibfnamefont {J.~M.}\ \bibnamefont {Abendroth}},\ }\bibfield  {title}
  {\bibinfo {title} {Diamond surface engineering for molecular sensing with
  nitrogen-vacancy centers},\ }\href {https://doi.org/10.1039/D2TC01258H}
  {\bibfield  {journal} {\bibinfo  {journal} {Journal of Materials Chemistry
  C}\ }\textbf {\bibinfo {volume} {10}},\ \bibinfo {pages} {13533} (\bibinfo
  {year} {2022})}\BibitemShut {NoStop}%
\bibitem [{\citenamefont {Petrini}\ \emph {et~al.}(2022)\citenamefont
  {Petrini}, \citenamefont {Tomagra}, \citenamefont {Bernardi}, \citenamefont
  {Moreva}, \citenamefont {Traina}, \citenamefont {Marcantoni}, \citenamefont
  {Picollo}, \citenamefont {Kvaková}, \citenamefont {Cígler}, \citenamefont
  {Degiovanni}, \citenamefont {Carabelli},\ and\ \citenamefont
  {Genovese}}]{petrini22}%
  \BibitemOpen
  \bibfield  {author} {\bibinfo {author} {\bibfnamefont {G.}~\bibnamefont
  {Petrini}}, \bibinfo {author} {\bibfnamefont {G.}~\bibnamefont {Tomagra}},
  \bibinfo {author} {\bibfnamefont {E.}~\bibnamefont {Bernardi}}, \bibinfo
  {author} {\bibfnamefont {E.}~\bibnamefont {Moreva}}, \bibinfo {author}
  {\bibfnamefont {P.}~\bibnamefont {Traina}}, \bibinfo {author} {\bibfnamefont
  {A.}~\bibnamefont {Marcantoni}}, \bibinfo {author} {\bibfnamefont
  {F.}~\bibnamefont {Picollo}}, \bibinfo {author} {\bibfnamefont
  {K.}~\bibnamefont {Kvaková}}, \bibinfo {author} {\bibfnamefont
  {P.}~\bibnamefont {Cígler}}, \bibinfo {author} {\bibfnamefont {I.~P.}\
  \bibnamefont {Degiovanni}}, \bibinfo {author} {\bibfnamefont
  {V.}~\bibnamefont {Carabelli}},\ and\ \bibinfo {author} {\bibfnamefont
  {M.}~\bibnamefont {Genovese}},\ }\bibfield  {title} {\bibinfo {title}
  {Nanodiamond–quantum sensors reveal temperature variation associated to
  hippocampal neurons firing},\ }\bibfield  {journal} {\bibinfo  {journal}
  {Advanced Science}\ }\textbf {\bibinfo {volume} {9}},\ \href
  {https://doi.org/10.1002/advs.202202014} {10.1002/advs.202202014} (\bibinfo
  {year} {2022})\BibitemShut {NoStop}%
\bibitem [{\citenamefont {Finco}\ and\ \citenamefont
  {Jacques}(2023)}]{finco23}%
  \BibitemOpen
  \bibfield  {author} {\bibinfo {author} {\bibfnamefont {A.}~\bibnamefont
  {Finco}}\ and\ \bibinfo {author} {\bibfnamefont {V.}~\bibnamefont
  {Jacques}},\ }\bibfield  {title} {\bibinfo {title} {Single spin magnetometry
  and relaxometry applied to antiferromagnetic materials},\ }\href
  {https://doi.org/10.1063/5.0167480} {\bibfield  {journal} {\bibinfo
  {journal} {APL Materials}\ }\textbf {\bibinfo {volume} {11}},\ \bibinfo
  {pages} {100901} (\bibinfo {year} {2023})}\BibitemShut {NoStop}%
\bibitem [{\citenamefont {Rovny}\ \emph {et~al.}(2024)\citenamefont {Rovny},
  \citenamefont {Gopalakrishnan}, \citenamefont {Jayich}, \citenamefont
  {Maletinsky}, \citenamefont {Demler},\ and\ \citenamefont
  {d.~Leon}}]{rovny24}%
  \BibitemOpen
  \bibfield  {author} {\bibinfo {author} {\bibfnamefont {J.}~\bibnamefont
  {Rovny}}, \bibinfo {author} {\bibfnamefont {S.}~\bibnamefont
  {Gopalakrishnan}}, \bibinfo {author} {\bibfnamefont {A.~C.~B.}\ \bibnamefont
  {Jayich}}, \bibinfo {author} {\bibfnamefont {P.}~\bibnamefont {Maletinsky}},
  \bibinfo {author} {\bibfnamefont {E.}~\bibnamefont {Demler}},\ and\ \bibinfo
  {author} {\bibfnamefont {N.~P.}\ \bibnamefont {d.~Leon}},\ }\bibfield
  {title} {\bibinfo {title} {New opportunities in condensed matter physics for
  nanoscale quantum sensors},\ }\href {https://arxiv.org/abs/2403.13710}
  {\bibfield  {journal} {\bibinfo  {journal} {arXiv:2403.13710}\ } (\bibinfo
  {year} {2024})}\BibitemShut {NoStop}%
\bibitem [{\citenamefont {der Sar}\ \emph {et~al.}(2015)\citenamefont {der
  Sar}, \citenamefont {Casola}, \citenamefont {Walsworth},\ and\ \citenamefont
  {Yacoby}}]{vandersar15}%
  \BibitemOpen
  \bibfield  {author} {\bibinfo {author} {\bibfnamefont {T.~V.}\ \bibnamefont
  {der Sar}}, \bibinfo {author} {\bibfnamefont {F.}~\bibnamefont {Casola}},
  \bibinfo {author} {\bibfnamefont {R.}~\bibnamefont {Walsworth}},\ and\
  \bibinfo {author} {\bibfnamefont {A.}~\bibnamefont {Yacoby}},\ }\bibfield
  {title} {\bibinfo {title} {Nanometre-scale probing of spin waves using single
  electron spins},\ }\href {https://doi.org/10.1038/ncomms8886} {\bibfield
  {journal} {\bibinfo  {journal} {Nat. Commun.}\ }\textbf {\bibinfo {volume}
  {6}},\ \bibinfo {pages} {7886} (\bibinfo {year} {2015})}\BibitemShut
  {NoStop}%
\bibitem [{\citenamefont {Wolfe}\ \emph {et~al.}(2016)\citenamefont {Wolfe},
  \citenamefont {Manuilov}, \citenamefont {Purser}, \citenamefont
  {Teeling-Smith}, \citenamefont {Dubs}, \citenamefont {Hammel},\ and\
  \citenamefont {Bhallamudi}}]{wolfe16}%
  \BibitemOpen
  \bibfield  {author} {\bibinfo {author} {\bibfnamefont {C.~S.}\ \bibnamefont
  {Wolfe}}, \bibinfo {author} {\bibfnamefont {S.~A.}\ \bibnamefont {Manuilov}},
  \bibinfo {author} {\bibfnamefont {C.~M.}\ \bibnamefont {Purser}}, \bibinfo
  {author} {\bibfnamefont {R.}~\bibnamefont {Teeling-Smith}}, \bibinfo {author}
  {\bibfnamefont {C.}~\bibnamefont {Dubs}}, \bibinfo {author} {\bibfnamefont
  {P.~C.}\ \bibnamefont {Hammel}},\ and\ \bibinfo {author} {\bibfnamefont
  {V.~P.}\ \bibnamefont {Bhallamudi}},\ }\bibfield  {title} {\bibinfo {title}
  {Spatially resolved detection of complex ferromagnetic dynamics using
  optically detected nitrogen-vacancy spins},\ }\href
  {https://doi.org/10.1063/1.4953108} {\bibfield  {journal} {\bibinfo
  {journal} {Appl. Phys. Lett.}\ }\textbf {\bibinfo {volume} {108}},\ \bibinfo
  {pages} {232409} (\bibinfo {year} {2016})}\BibitemShut {NoStop}%
\bibitem [{\citenamefont {Zhou}\ \emph {et~al.}(2020)\citenamefont {Zhou},
  \citenamefont {Jerger}, \citenamefont {Lee}, \citenamefont {Fukami},
  \citenamefont {Mujid}, \citenamefont {Park},\ and\ \citenamefont
  {Awschalom}}]{zhou20}%
  \BibitemOpen
  \bibfield  {author} {\bibinfo {author} {\bibfnamefont {B.~B.}\ \bibnamefont
  {Zhou}}, \bibinfo {author} {\bibfnamefont {P.~C.}\ \bibnamefont {Jerger}},
  \bibinfo {author} {\bibfnamefont {K.}~\bibnamefont {Lee}}, \bibinfo {author}
  {\bibfnamefont {M.}~\bibnamefont {Fukami}}, \bibinfo {author} {\bibfnamefont
  {F.}~\bibnamefont {Mujid}}, \bibinfo {author} {\bibfnamefont
  {J.}~\bibnamefont {Park}},\ and\ \bibinfo {author} {\bibfnamefont {D.~D.}\
  \bibnamefont {Awschalom}},\ }\bibfield  {title} {\bibinfo {title}
  {Spatiotemporal mapping of a photocurrent vortex in monolayer
  ${\mathrm{mos}}_{2}$ using diamond quantum sensors},\ }\href
  {https://doi.org/10.1103/PhysRevX.10.011003} {\bibfield  {journal} {\bibinfo
  {journal} {Phys. Rev. X}\ }\textbf {\bibinfo {volume} {10}},\ \bibinfo
  {pages} {011003} (\bibinfo {year} {2020})}\BibitemShut {NoStop}%
\bibitem [{\citenamefont {Bloch}\ and\ \citenamefont
  {Siegert}(1940)}]{bloch40}%
  \BibitemOpen
  \bibfield  {author} {\bibinfo {author} {\bibfnamefont {F.}~\bibnamefont
  {Bloch}}\ and\ \bibinfo {author} {\bibfnamefont {A.}~\bibnamefont
  {Siegert}},\ }\bibfield  {title} {\bibinfo {title} {Magnetic resonance for
  nonrotating fields},\ }\href {https://doi.org/10.1103/PhysRev.57.522}
  {\bibfield  {journal} {\bibinfo  {journal} {Phys. Rev.}\ }\textbf {\bibinfo
  {volume} {57}},\ \bibinfo {pages} {522} (\bibinfo {year} {1940})}\BibitemShut
  {NoStop}%
\bibitem [{\citenamefont {Fuchs}\ \emph {et~al.}(2009)\citenamefont {Fuchs},
  \citenamefont {Dobrovitski}, \citenamefont {Toyli}, \citenamefont
  {Heremans},\ and\ \citenamefont {Awschalom}}]{fuchs09}%
  \BibitemOpen
  \bibfield  {author} {\bibinfo {author} {\bibfnamefont {G.~D.}\ \bibnamefont
  {Fuchs}}, \bibinfo {author} {\bibfnamefont {V.~V.}\ \bibnamefont
  {Dobrovitski}}, \bibinfo {author} {\bibfnamefont {D.~M.}\ \bibnamefont
  {Toyli}}, \bibinfo {author} {\bibfnamefont {F.~J.}\ \bibnamefont
  {Heremans}},\ and\ \bibinfo {author} {\bibfnamefont {D.~D.}\ \bibnamefont
  {Awschalom}},\ }\bibfield  {title} {\bibinfo {title} {Gigahertz dynamics of a
  strongly driven single quantum spin},\ }\href
  {https://doi.org/10.1126/science.1181193} {\bibfield  {journal} {\bibinfo
  {journal} {Science}\ }\textbf {\bibinfo {volume} {326}},\ \bibinfo {pages}
  {1520} (\bibinfo {year} {2009})}\BibitemShut {NoStop}%
\bibitem [{\citenamefont {Kairys}\ \emph {et~al.}(2023)\citenamefont {Kairys},
  \citenamefont {Marcks}, \citenamefont {Delegan}, \citenamefont {Zhang},
  \citenamefont {Awschalom},\ and\ \citenamefont {Heremans}}]{kairys23}%
  \BibitemOpen
  \bibfield  {author} {\bibinfo {author} {\bibfnamefont {P.}~\bibnamefont
  {Kairys}}, \bibinfo {author} {\bibfnamefont {J.~C.}\ \bibnamefont {Marcks}},
  \bibinfo {author} {\bibfnamefont {N.}~\bibnamefont {Delegan}}, \bibinfo
  {author} {\bibfnamefont {J.}~\bibnamefont {Zhang}}, \bibinfo {author}
  {\bibfnamefont {D.~D.}\ \bibnamefont {Awschalom}},\ and\ \bibinfo {author}
  {\bibfnamefont {F.~J.}\ \bibnamefont {Heremans}},\ }\bibfield  {title}
  {\bibinfo {title} {Quantifying the limits of controllability for the
  nitrogen-vacancy electron spin defect},\ }\href
  {https://arxiv.org/abs/2309.03120} {\bibfield  {journal} {\bibinfo  {journal}
  {arXiv:2309.03120}\ } (\bibinfo {year} {2023})}\BibitemShut {NoStop}%
\bibitem [{\citenamefont {Baumgartner}\ \emph {et~al.}(2017)\citenamefont
  {Baumgartner}, \citenamefont {Garello}, \citenamefont {Mendil}, \citenamefont
  {Avci}, \citenamefont {Grimaldi}, \citenamefont {Murer}, \citenamefont
  {Feng}, \citenamefont {Gabureac}, \citenamefont {Stamm}, \citenamefont
  {Acremann}, \citenamefont {Finizio}, \citenamefont {Wintz}, \citenamefont
  {Raabe},\ and\ \citenamefont {Gambardella}}]{baumgartner17}%
  \BibitemOpen
  \bibfield  {author} {\bibinfo {author} {\bibfnamefont {M.}~\bibnamefont
  {Baumgartner}}, \bibinfo {author} {\bibfnamefont {K.}~\bibnamefont
  {Garello}}, \bibinfo {author} {\bibfnamefont {J.}~\bibnamefont {Mendil}},
  \bibinfo {author} {\bibfnamefont {C.~O.}\ \bibnamefont {Avci}}, \bibinfo
  {author} {\bibfnamefont {E.}~\bibnamefont {Grimaldi}}, \bibinfo {author}
  {\bibfnamefont {C.}~\bibnamefont {Murer}}, \bibinfo {author} {\bibfnamefont
  {J.}~\bibnamefont {Feng}}, \bibinfo {author} {\bibfnamefont {M.}~\bibnamefont
  {Gabureac}}, \bibinfo {author} {\bibfnamefont {C.}~\bibnamefont {Stamm}},
  \bibinfo {author} {\bibfnamefont {Y.}~\bibnamefont {Acremann}}, \bibinfo
  {author} {\bibfnamefont {S.}~\bibnamefont {Finizio}}, \bibinfo {author}
  {\bibfnamefont {S.}~\bibnamefont {Wintz}}, \bibinfo {author} {\bibfnamefont
  {J.}~\bibnamefont {Raabe}},\ and\ \bibinfo {author} {\bibfnamefont
  {P.}~\bibnamefont {Gambardella}},\ }\bibfield  {title} {\bibinfo {title}
  {Spatially and time-resolved magnetization dynamics driven by spin-orbit
  torques},\ }\href {https://doi.org/10.1038/nnano.2017.151} {\bibfield
  {journal} {\bibinfo  {journal} {Nature Nanotechnology}\ }\textbf {\bibinfo
  {volume} {12}},\ \bibinfo {pages} {980} (\bibinfo {year} {2017})}\BibitemShut
  {NoStop}%
\bibitem [{\citenamefont {Herb}\ \emph {et~al.}(2024)\citenamefont {Herb},
  \citenamefont {Völker}, \citenamefont {Abendroth}, \citenamefont
  {Meinhardt}, \citenamefont {van Schie}, \citenamefont {Gambardella},\ and\
  \citenamefont {Degen}}]{arxiv_nanosecond}%
  \BibitemOpen
  \bibfield  {author} {\bibinfo {author} {\bibfnamefont {K.}~\bibnamefont
  {Herb}}, \bibinfo {author} {\bibfnamefont {L.~A.}\ \bibnamefont {Völker}},
  \bibinfo {author} {\bibfnamefont {J.~M.}\ \bibnamefont {Abendroth}}, \bibinfo
  {author} {\bibfnamefont {N.}~\bibnamefont {Meinhardt}}, \bibinfo {author}
  {\bibfnamefont {L.}~\bibnamefont {van Schie}}, \bibinfo {author}
  {\bibfnamefont {P.}~\bibnamefont {Gambardella}},\ and\ \bibinfo {author}
  {\bibfnamefont {C.~L.}\ \bibnamefont {Degen}},\ }\href
  {https://doi.org/10.48550/arXiv.2411.05542} {\bibinfo {title} {Quantum
  magnetometry of transient signals with a time resolution of 1.1 nanoseconds}}
  (\bibinfo {year} {2024}),\ \Eprint {https://arxiv.org/abs/2411.05542}
  {arXiv:2411.05542 [quant-ph]} \BibitemShut {NoStop}%
\end{thebibliography}%

\end{document}


\large
\begin{center}
\textbf{{Supplemental material for: \\ ``Quantum speed limit in quantum sensing''}}

\normalsize

\vspace{10 mm}

Konstantin~Herb and Christian~L.~Degen

\vspace{10 mm}

\end{center}
\small

\FloatBarrier

\section{Bipartite control sequence}
We derive an analytical expression for the transition probability [Eq.~5] for the bi-partite control sequence of Fig.~1(e). In the rotating frame, the Hamiltonian of the NV center is given by
\begin{equation}
	\HH/\hbar = \dw \Sz + \Omega(t) \left(\Sy\cos\theta  + \Sx\sin\theta \right) ,
\end{equation}
where $\dw$ is the detuning of the NV center from the resonance frequency (\ie, the stimulus to be detected), $\Omega(t)$ is the shaped amplitude of control pulses, $\theta$ is the phase of control pulses, and $\Si$ ($i=x,y,z$) are spin-1/2 matrices.  We assume that the control pulses induce a rotation angle $\alpha$ during the time $\tRR = \tau/2$ with $\theta = 0$, followed by a pulse of length $\tRRd=\tau/2$ with $\theta = \pi/2$.  The Hamiltonian is piece-wise constant for a square amplitude, $\Omega(t) = \Omega$. Therefore, we can analytically calculate the propagator $\UU$. For an NV center initialized and detected in $\ket{0}$, the transition probability $p$ is given by
\begin{equation}
	p = 1 - \left|\langle 0 | \UU | 0 \rangle\right|^2 .
\end{equation}
For the above control pulses, $p$ is given through
\begin{align}
	1-p= \frac{\Omega ^2 \left(4 \dw ^2 \cos \left(\frac{\tau \tilde{\Omega }}{2}\right)+8 \dw  \tilde{\Omega
		} \sin ^2\left(\frac{\tau \tilde{\Omega }}{4}\right) \sin \left(\frac{\tau \tilde{\Omega }}{2}\right)+\Omega ^2 \cos
		\left(\tau \tilde{\Omega }\right)\right)+4 \dw^4+4 \dw^2 \Omega ^2+3 \Omega ^4}{4 \tilde{\Omega
		}^4} ,
\end{align}
where $\tilde{\Omega } = \sqrt{\Omega^2+\Delta\omega^2}$ is the effective Rabi frequency.

Practically, we will only consider situations where the control pulses are much stronger than the stimulus to be detected, that is, $\dw \ll \Omega$.  Therefore, we can expand $p$ to first order in $\dw/\Omega$.  We find
\begin{align}
	p &= \frac{1}{4}-\frac{1}{4} \cos (\Omega \tau)
	+ \frac{\dw}{\Omega}\sin \left(\frac{\Omega \tau}{2}\right) \left( \cos \left(\frac{\Omega \tau}{2}\right) - 1\right)  \\
    &= \frac{1}{4}\left(1-\cos 2\alpha\right)
    + \frac{\phi}{2\alpha}\sin\alpha \left(\cos\alpha - 1\right)
    \equiv \text{Eq. (6)} \label{eq:p} ,
\end{align}
where $\alpha=\Omega\tau/2$ is the rotation angle of each pulse and $\phi=\dw\tau$ is the Ramsey phase.  Eq.~\ref{eq:p} defines an effective phase given by
\begin{align}
	\phieff &= \frac{\sin\alpha(\cos\alpha-1)\dw\tau}{\alpha} ,
\end{align}
see solid curve in Fig.~2.

Eq.~\ref{eq:p} applies to the control sequence shown in Fig.~1(e). 
We can extend Eq.~\ref{eq:p} to the more general case of Fig.~1(d), that is, to a control sequence including a free evolution period.  For this, we assume that $\alpha = 90^\circ$ and that $\tau>2\tRR$ is longer than the control rotations.  Then, the total pulse duration is $2\tRR$, and the free evolution period is $\tau - 2\tRR$.  The total accumulated phase is given by the sum of the phases accumulated during control rotations and during free evolution,
\begin{align}
	\phieff &= \frac{4\dw\tRR}{\pi} + \dw(\tau-2\tRR) ,
\end{align}
see dashed curve in Fig.~2.  The corresponding transition probability is
\begin{align}
	p &= \frac{1}{2} + \frac{2\dw\tRR}{\pi} + \frac{\dw (\tau-2\tRR)}{2} .
\end{align}

\FloatBarrier
\section{Optimal choice of control rotations of the bipartite control sequence}
\FloatBarrier
This section aims to show that for a two-partite pulse, the choice of equal timeshare and a phase jump of 90\textdegree is the optimal choice.  By optimal choice, we mean that the sensitivity
\begin{equation}
	\eta = \left.\frac{\partial [\delta p]}{\partial [\dw]}\right\vert_{\dw \rightarrow 0}
\end{equation}
is maximal in its magnitude. We fix the pulse length and allow the Rabi frequency to vary to achieve a given timing resolution. The choice remains optimal if the Rabi frequency is limited (and therefore the maximal achievable flip angle $\alpha$).

For this, we calculate the propagator under a bipartite Hamtilonian
\begin{align}
	\hat{H}_1 &= \Omega \hat{S}_y + \dw \hat{S}_z \\
	\hat{H}_2 &= \Omega \left(\hat{S}_y  \cos\theta + \hat{S}_x \sin\theta \right) + \dw \hat{S}_z
\end{align}
We let the system evolve for the time $k\tau$ under $\hat{H}_1$ and afterwards for $(1-k)\tau$ under $\hat{H}_2$. From this, we compute the transition probability
\begin{equation}
	1-p = \left|\left\langle0 \left| e^{-\ii \hat{H}_2 (1-k) \tau} e^{-\ii \hat{H}_1 k \tau} \right| 0\right\rangle\right|^2
\end{equation}
This allows us to find an analytical expression for the sensitivity 
\begin{equation}
	\eta = \sin\theta \frac{ \sin( (k-1) \Omega \tau)-\sin(k  \Omega \tau)+\sin (\Omega\tau )}{2 \Omega }
\end{equation}
With regard to the angle $\theta$, we see that the sensitivity is maximized for $\theta=\pi/2$, thus for doing the second pulse segment along an orthogonal axis. For the optimal choice of $k$, we find the root of the derivative $\partial S/\partial k$ yielding
\begin{equation}
	\cos{\left((1-k) \,\Omega \tau\right)} = \cos{\left(k\, \Omega \tau\right)}
\end{equation}
and thus an optimal choice of $k=1/2$.

\clearpage
\FloatBarrier

\section{Extended series of Bloch sphere representations of the qubit evolution}

\begin{figure}[h!]
	\centering
	\includegraphics[width=0.98\textwidth]{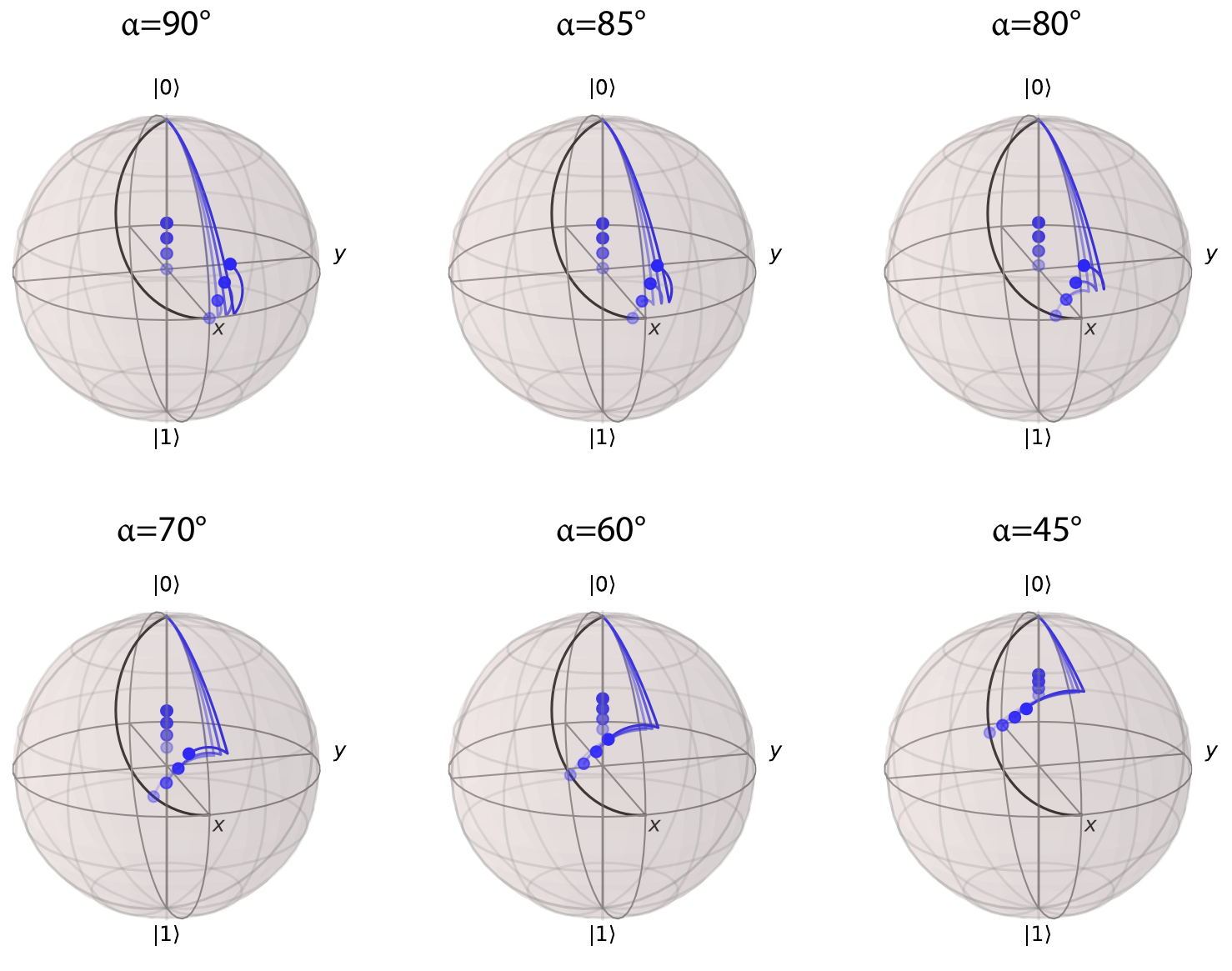}
	\caption{Series of qubit trajectories for decreasing angles $\alpha = 90^\circ$ to $\alpha=45^\circ$, complementing Fig.~1(f,g) in the main text.  Dots are the projection on the $Z$ axis. 	Trajectories are shown for $\phi = 0^\circ$ (light blue), $10^\circ$, $20^\circ$ and $30^\circ$ (dark blue).  Black curves on all spheres show the trajectory endpoints for $\phi=0$ as a function of $\alpha=0...90^\circ$.}
\end{figure}

\clearpage
\FloatBarrier

\section{Definition of time resolution and bandwidth}

In this section, we give alternative definitions for time resolution and bandwidth of the bi-partite control sequence.

\subsection{Time resolution}
\textit{Full width at half maximum} -- 
A common choice used in the main text is quantifying the time resolution by the full width at half maximum (FWHM) of the kernel function.  This definition yields
\begin{align}
	t_\mathrm{min} ={\tau} - \frac{2}{\Omega} \arcsin \frac{\sin \left(\frac{\tau}{2\Omega}\right)}{2} = \tau \left(1- \frac{\arcsin\frac{\sin\alpha}{2}}{\alpha}\right),
\end{align}
where $\alpha = \Omega\tau/2$.  For $\alpha=\pi/2$, this is
\begin{align}
	t_\mathrm{min}^{(\pi/2)} = \frac{2}{3}\tau \approx 0.667\tau.
\end{align}

\textit{Rise time} -- 
Another definition is through the 10-90\% or 20-80\% rise time, \ie, the time required for the kernel function to rise from low to high amplitude.  These definitions yield
\begin{align}
	t_{20-80}
	&= \tau - \frac{2}{\Omega} \arccos\left(\frac{2}{5}\cos\left(\frac{\Omega \tau}{2}\right) + \frac{3}{5}\right)  \ ,  \\
	t_{10-90}
	&= \tau - \frac{2}{\Omega} \arccos\left(\frac{1}{5}\cos\left(\frac{\Omega \tau}{2}\right) + \frac{4}{5}\right),
\end{align}
respectively. For $\alpha=\pi/2c$, these become
\begin{align}
	t_{20-80}^{(\pi/2)}
	&= \tau \left(1-\frac{1}{\pi}\arccos\frac{1}{5}\right) \approx 0.564 \tau \ , \\
	t_{10-90}^{(\pi/2)}
	&= \tau \left(1-\frac{1}{\pi}\arccos\frac{3}{5}\right) \approx 0.704 \tau .
\end{align}

\textit{Equivalent duration} --
Another common choice is the equivalent duration of a square kernel with the same height as the peak height of the kernel and an equivalent area. The corresponding time resolution is
\begin{align}
	t_{\scriptscriptstyle\square}
	&= \frac{1}{k(0)}\int_{-\tau/2}^{\tau/2} k(t) dt = 
	\frac{1}{\sin(\Omega\tau/2)} \frac{4 \sin^2 (\Omega\tau/4)}{\Omega}
	= \frac{2}{\Omega} \tan\left(\frac{\Omega\tau}{4}\right)
	= \tau \frac{\tan\frac{\alpha}{2}}{\alpha}
\end{align}
For $\alpha=\pi/2$, this is
\begin{align}
	t_{\scriptscriptstyle\square}^{(\pi/2)} = \frac{2}{\pi}\tau \approx 0.637\tau
\end{align}
\pagebreak

\subsection{Bandwidth}

Another important figure of merit is the instantaneous bandwidth, roughly given by the inversely of the time resolution.

\textit{First root of the kernel spectrum} --
A standard definition is by the first root of the kernel transfer function $K(\omega)$, which reflects the maximum spectral component that the sequence can keep pace with.  Applying this definition,
\begin{align}
	\BW = \Omega \left(\frac{2\pi}{\alpha}-1\right)
\end{align}
For $\alpha=\pi/2$, this is
\begin{align}
	\BW^{(\pi/2)} = 3\Omega.
\end{align}

\textit{3-dB point} --
Another common definition is \textit{via} the 3-dB point, defined by $K_\mr{3dB}(\omega) := \frac{1}{\sqrt{2}} K(0)$. Taking {Eq.~(12)}, the following transcendental equation has to be {numerically} solved for $y$:
\begin{align}
	\frac{y^2-1}{\sqrt{2}}\left({1-\cos\alpha}\right) = \left|\cos\alpha-\cos y\alpha\right|
\end{align}
Then, the bandwidth is given by 
\begin{align}
	\Omega_{3\,\mathrm{dB}} = y \,\Omega.
\end{align}
For $\alpha=\pi/2$, the solution is
\begin{align}
	\Omega_{3\,\mathrm{dB}}^{(\pi/2)} \approx 1.19\, \Omega.
\end{align}
\FloatBarrier

\section{Spin simulations}
\label{sec:simulations}

To investigate the limit of an NV probe qutrit, we perform laboratory frame simulations with an explicitly time-dependent Hamiltonian.
\begin{equation}
	\hat{H} = D \hat{S}_z^2  + \gamma_e B_0 \hat{S}_z + \gamma_e B_1 m(t) \hat{S}_x + \gamma_e B_\mathrm{stim}(t) S_z
	\label{eq:hamiltonian}
\end{equation}
where $\gamma_e=28.0345\unit{GHz/T}$ denotes the gyromagnetic ratio of the NV center and $D=2\pi\cdot2.87\unit{GHz}$ the zero-field splitting. $B_1$ denotes the magnitude of the driving RF field that performs the qutrit rotations, whereas $m(t)$ denotes the modulation at the driven resonance frequency; thus, it is a sinusoidal wave at frequency $\omega$ [c.f. Fig. 4(a)]. The Rabi frequency is $\Omega=\gamma_\mathrm{e} B_1 / \sqrt{2}$. This is because the Rabi frequency of a spin-1 system is $\sqrt{2}$ times larger than for a spin-1/2. The transient to sense is in general the time-dependent field $B_\mathrm{stim}(t)$.

To investigate the limits to time resolution, we need to shorten the control pulses and, therefore, increase $B_1$. In the simulation, we keep the flip angle fixed at 90\textdegree. Stronger and stronger control pulses will eventually lead to strong driving effects due to the breakdown of the rotating wave approximation (RWA). This can be avoided by increasing $B_0$, ensuring the RWA is fulfilled. It is vital to notice that this does not mitigate the saturation of the detuning of the two transition frequencies $\omega$ and $\omega'$ as shown in Fig. 4(c). To obtain the Fig. 4(d), we selected a bias field of $B_0=40\unit{T}$. The stimulus is a constant offset where we reverse the direction in a second simulation run to obtain the upper and the lower line in Fig. 4(d). The stimulus signal is increased to ensure a constant phase pickup and selected to be $B_\mathrm{stim}=B_1/(10\sqrt{2})$. This yields an effective phase pickup of $\phieff=1/5$ constant for all pulse durations. %

\FloatBarrier
\section{Off-axis effects}
\FloatBarrier
To simulate off-axis effects, we perform laboratory frame simulations using the framework of Section~\ref{sec:simulations}. %
We modify the Hamiltonian in Eq. \eqref{eq:hamiltonian} to allow for stimuli along an arbitrary axis.  For this purpose, we introduce the angle $\chi$, which describes the tilting of the field with respect to the qubit quantization axis (by convention, the $Z$ axis): %
\begin{equation}
	\hat{H} = D \hat{S}_z^2  + \gamma_\mathrm{e} B_0 \hat{S}_z + \gamma_\mathrm{e} B_1 m(t) \hat{S}_x + \gamma_\mathrm{e} B_\mathrm{stim}(t) \left(\cos(\chi) S_z + \sin(\chi) S_x \right)
	\label{eq:hamiltonian2}
\end{equation}
To obtain the sensing kernel, we propagate numerically a Gaussian with an FWHM that is approximately ten times smaller than the expected kernel width. %
To obtain the Bode plot, we numerically apply a sinusoidal RF tone as $B_\mathrm{stim}(t)$ at increasing frequencies. We then fit the amplitude of the propagated sine wave to obtain the Bode plot. This allows examining the frequency behavior more carefully. We observe that for RF frequencies $\lesssim 2.5 \Omega$, we do not observe any significant difference to the fully aligned case. 

Furthermore, we show for a $\chi=45$\textdegree~angle the entire plot up to the Larmor frequency. We see that when approaching the Larmor frequency, the role of the off-axis component becomes increasingly important. %

\begin{figure}[h!]
	\centering
	\includegraphics[width=0.9\textwidth]{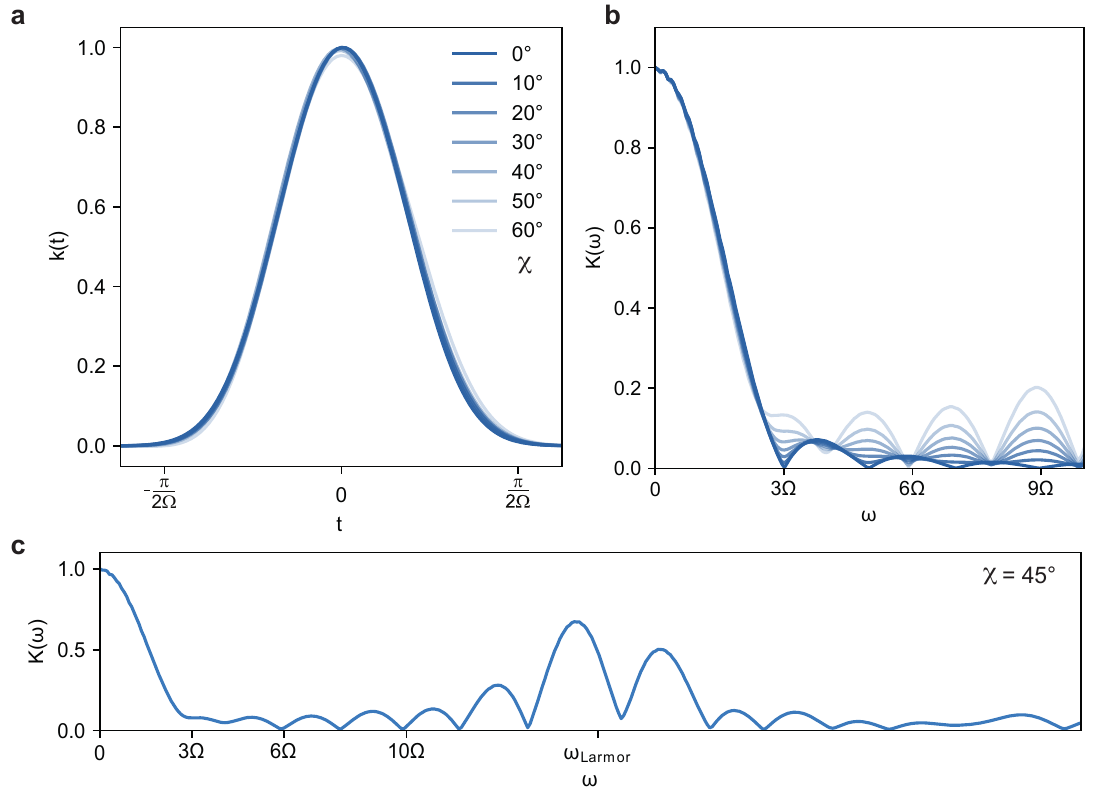}
	\caption{Effect of off-axis transients. \textbf{a} Numerically estimated sensing kernel for various off-axis angles $\chi$. \textbf{b} Corresponding Bode plots for the frequency range shown in the main text. \textbf{c} Bode plot for $\chi=45$\textdegree for a larger frequency range.}
\end{figure}